\documentclass[showpacs,superscriptaddress,notitlepage,prb,onecolumn,longbibliography]{revtex4-1}
\usepackage[pdftex]{graphicx}
\usepackage{amsfonts}
\usepackage{mathtools}
\usepackage{amssymb}
\usepackage{upgreek}
\usepackage{dsfont}
\usepackage{bm}
\usepackage{color}
\usepackage[colorlinks=true]{hyperref}
\usepackage{multirow}
\usepackage{pifont}
\usepackage{float}
\usepackage[caption = false]{subfig}
\newcommand{\cmark}{\ding{51}}%
\newcommand{\xmark}{\ding{55}}

\begin{document}

\title{Pseudospin-valve effect on transport in junctions of three-dimensional topological insulator surfaces}
\author{Sthitadhi Roy}
\affiliation{Max-Planck-Institut f\"ur Physik komplexer Systeme, N\"othnitzer Stra$\beta$e 38, 01187 Dresden, Germany}
\author{Krishanu Roychowdhury}
\affiliation{Max-Planck-Institut f\"ur Physik komplexer Systeme, N\"othnitzer Stra$\beta$e 38, 01187 Dresden, Germany}
\affiliation{LASSP, Department of Physics, Cornell University, Ithaca, NY 14853, USA}
\author{Sourin Das}
\affiliation{Department of Physics and Astrophysics, University of Delhi, Delhi 110 007, India}

\date{\today}
\begin{abstract}
We show that the surface states of pristine 3D topological insulators (TI) are analogs of ferromagnetic half metals due to complete polarization of an emergent  momentum independent pseudospin (SU(2)) degree of freedom on the surface. To put this claim on firm footing, we present results for TI surfaces perpendicular to the crystal growth axis, which clearly show that the tunneling conductance between two such TI surfaces of the same TI material is dominated by this half metallic behavior leading to physics reminiscent of a spin-valve. Further using the generalized tunnel magnetoresistance derived in this work we also study the tunneling current between arbitrary TI surfaces. We also perform a comprehensive study of the effect of all possible surface potentials allowed by time reversal symmetry on this spin-valve effect and show that it is robust against most of such potentials.
\end{abstract}
\maketitle
\section{Introduction \label {sec:intro}}

Three dimensional topological insulators (3D TI) \cite{Fu2007, Moore2010, Hasan2010, Qi2011,Zhang2009} represent a distinct class of 3D band insulators which have  topologically protected metallic surface states. Nontrivial topology of band structures of these materials leads to the immunity of these surface states against a variety of disorder and interaction potentials. Possibility of surface magnetization and related spin textures of Fermi surface in these materials due to external doping\cite{Chen2010,Zhu2011,Wray2011,Henk2012,Xu2012,Checkelsky2012} has also been a topic of great interest. Here we show that, even the pristine TI surface can act like a ferromagnetic half metal\cite{Groot83}, though not due to spin but due to an emergent pseudospin degree of freedom\cite{Zhang2012}.
A natural way to confirm a half-metallic behavior is an observation of perfect spin orthogonality (complete suppression of tunneling due to orthogonality of the states in the spin sector) in tunnel conductance between two half-metals with opposite directions of polarization and then a lifting of the orthogonality by tilting the polarization direction of one half-metal with respect to another. The central result of this work is the discovery of such (pseudo)spin orthogonality physics and its subsequent lifting  due to the rotation of pseudospin polarizations on the surface states of 3D TIs.

The low energy physics of popular TI materials like $\text{Bi}_2\text{Se}_3$ can be described by a four band model arising from two SU(2) degrees of freedom.\cite{Zhang2009,Liu2010,Zhang2012, Zhang2013}. Of these two SU(2)  degree of freedom,  one of them is dispersing \footnote{observables which have momentum dependent expectation value in eigenstate of surface Hamiltonian} while the other one is non-dispersing.  This fact is natural as the surface state itself comes into being by freezing (making it momentum independent) two out of four bulk  degrees of freedom (four band model) on the planar boundary. This makes the frozen SU(2)  degree of freedom fully polarized and hence surface state is endowed with a half-metal like character.

 It is interesting to note that the polarization of the frozen degree of freedom is exactly opposite on the opposite surfaces for any angle $\theta$\cite{Zhang2012,Roy2014a} that the surfaces make with the crystal growth axis, hence leading to distinction between the top and bottom surfaces which is very similar to the  distinction of north and south poles of a bar magnet. It can be understood from the observation that, when a bar magnet is broken into two halves, it exposes two new ends with opposite polarity such that each new bar magnet has opposite polarity at its two ends. In an identical fashion when a sample of TI with exposed surfaces corresponding to $\theta$ and $\theta+\pi$ is sliced parallel to the plane creating an angle $\theta$ with the crystal growth axis, creating two new TI films, the newly exposed surfaces will have exactly opposite polarizations in the frozen degree of freedom consistent with the scenario of a bar magnet. The fact that the different surfaces of the TI are very distinct from one another due to this frozen  degree of freedom has been mostly underestimated and ignored. One of the most important outcome of our analysis is, though frozen, these   degree of freedom can have strong influence on transport across junctions of TI via physics analogous to the spin-valve effect. In our analysis, the role of the edge states which may appear at the edges of a 3D sample\cite{Zhang2013,Deb2014} will have a negligible effect and hence are neglected.
 
 The rest of the paper is organized is follows, in Sec.\ref{sec:model} we describe the model Hamiltonian for the surface states their pseudospin textures. In Sec.\ref{sec:gtmr} we sketch the calculation of the tunneling current between two TI surfaces and derive the expression for the generalized tunnel magnetoresistance followed by a discussion of identification of the half-metallic behavior via tunneling current in Sec.\ref{sec:halfmetal}. In Sec.\ref{sec:arbitrary} we discuss the effect of interplay of the general tunnel magnetoresistance and the shape of Fermi surface on tunneling current between two arbitrary surfaces. The robustness of the half-metallic behavior against surface potentials is discussed in Sec.\ref{sec:surfacepotentials} followed by concluding remarks in Sec.\ref{sec:conclusions}.
 
\section{Model and preliminaries \label{sec:model}}
In this section, we describe the model Hamiltonian for the surface states of the 3D TI. We sketch the derivations of the surface state spinors and their properties, setting up consistent notations used throughout the rest of the paper. For details refer to Refs.[\onlinecite{Zhang2009,Liu2010,Zhang2012,Roy2014a}].
\begin{figure}
\includegraphics[width=0.5\columnwidth]{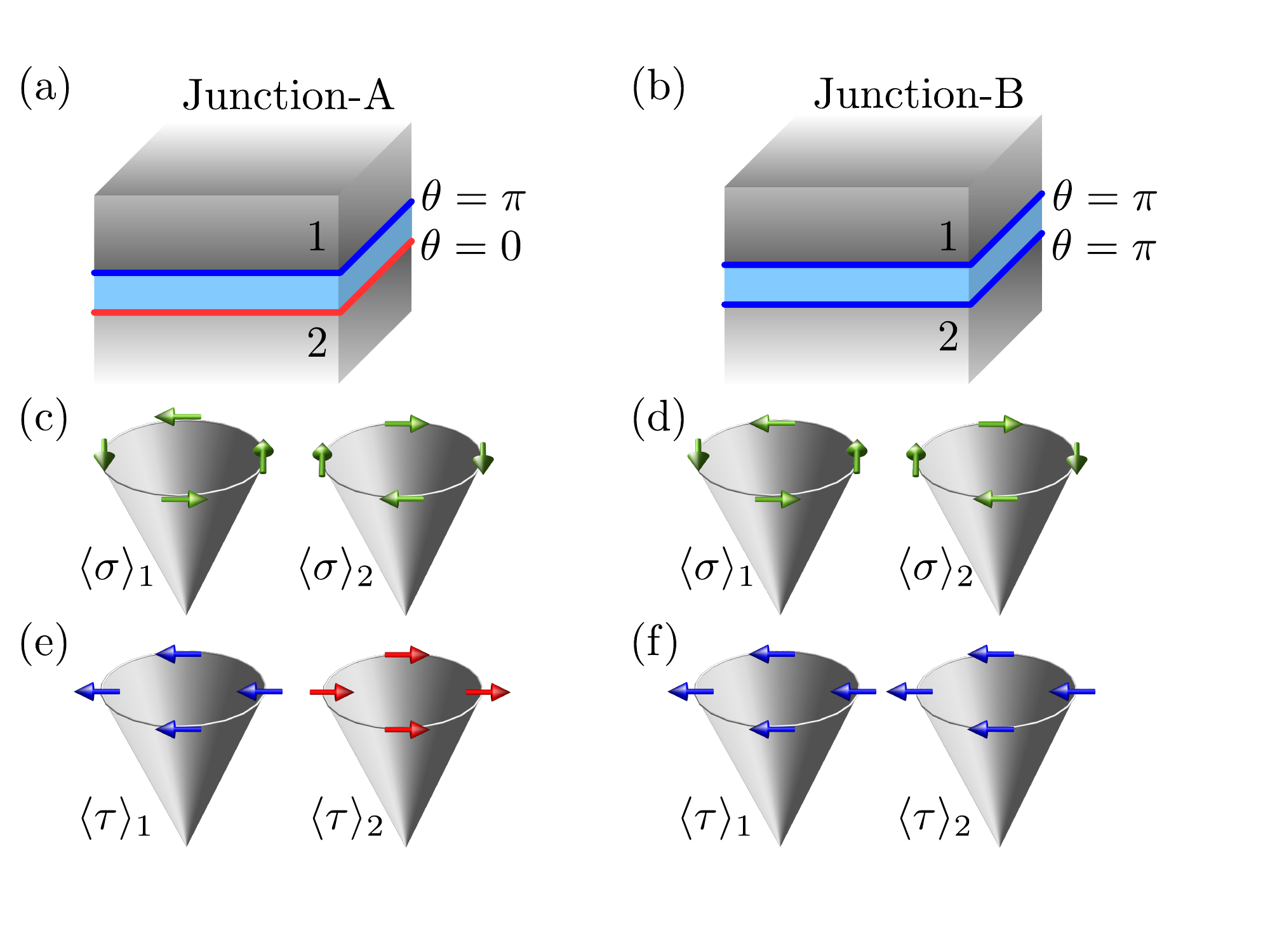}
\caption{(a) and (b): A schematic showing the two different junctions A and B of TI surfaces with an insulating barrier between them. The spin textures [(c) and (d)] and the orbital pseudospin textures [(e) and (f)] of the two surfaces are plotted on the Dirac cones of the surface states. The orbital pseudospin textures are completely polarized along the positive (blue) or negative (red) $x$ direction of the $\tau$-space.}
\label{fig:setup}
\end{figure}

Popular 3D TI material crystals like $\text{Bi}_2\text{Se}_3$ have a quintuple layered unit cell comprising of two equivalent Se layers (Se1 and Se1$^\prime$), two equivalent Bi layers (Bi1 and Bi1$^\prime$) and one inequivalent Se layer (Se2). Both, Bi and Se have the $p$-orbitals as their outermost shells, hence it is natural to consider only the $p$-orbitals for the electronic properties. As the hybridization of the Bi and Se orbitals, formation of the anti-bonding and bonding orbitals, crystal field splitting, and the effect of spin-orbit coupling are taken into account, it turns out that there are four states close to the Fermi level which participate in the low-energy description of the materials. We briefly discuss the effect of each of these steps in the following:
\begin{itemize}
\item The hybridization of the Bi and Se orbitals pushes up the Bi energy level forming new hybridized degenrate levels denoted by $B$ and $B^\prime$ , where as the Se energy levels are pushed down forming new hybridized levels denoted as $S$, $S^\prime$ and $S0$, out of which $S$ and $S^\prime$ are degenerate. 
\item Among the four states near the Fermi level two of them arise from the bonding orbital between the $B$ and $B^\prime$, denoted as $P1^+$, and the other two states arise from the anti-bonding orbital between $S$ and $S^\prime$ denoted as $P2^-$. Note that due to inversion symmetry, the bonding and anti-bonding orbitals have definite parity. 
\item Each of the two states coming from $P1^+$ or $P2^-$ have equal and opposite total angular momentum (sum of orbital angular momentum and spin) which turns out to be $\pm1/2$. The states with total angular momentum $+1/2$ are constituted by linear combination of $p_z$ orbitals with $\uparrow$ spin and $p_+=p_x+ip_y$ orbitals with $\downarrow$ spin. Similarly a linear combination of $p_z$ orbitals with $\downarrow$ spin and $p_-=p_x-ip_y$ orbitals with $\uparrow$ spin form the states with total angular momentum $-1/2$. 
\item Hence the four states near the Fermi energy can be represented as $\vert P1^+,\pm\frac{1}{2}\rangle$ and $\vert P2^-,\pm\frac{1}{2}\rangle$. Note that this naturally provides us with two SU(2) degrees of freedom to represent the states. The first SU(2) degree of freedom represented by $P1^+$ and $P2^-$ have different orbital parities, hence we call it the \emph{orbital pseudospin} and denote it by the set of Pauli matrices $\bm{\tau}$. It also turns out that for the surface states, the spin angular momentum is proportional to the total angular momentum\cite{Liu2010}, hence we call the other SU(2) degree of freedom, the total angular momentum simply the \emph{spin} and denote it by the set of Pauli matrices $\bm{\sigma}$.
\end{itemize}
As mentioned in Sec.\ref{sec:intro}, the surface states are labeled by an angle $\theta$ between the crystal growth axis and the normal to the surface. For the simple case of $\theta=0$ (top surface) and  $\theta=\pi$ (bottom surface), the dispersing and frozen  degree of freedom are given by $\bm{\sigma}$ and $\bm{\tau}$ operators respectively. The direction of spin, $\bm{\sigma}$, is determined via the spin-momentum locking angle while the $\bm{\tau}$ points solely along the positive $x$ (negative $x$) direction for the top (bottom) surface\cite{Zhang2012,Roy2014a} for all  momentum states hence defining a $\tau$ polarized half metal. \footnote{Note that the direction for $\tau$ vector lives in an abstract space which is not related to the physical space in which the crystal lives}. \\

\subsection{Surface states \label{sec:surfacestates}}

In a coordinate system, where the exposed TI surfaces lie parallel to the $x$-$y$ plane with its crystal growth axis at an angle $\theta$ to the $z$-axis, the composite Hamiltonian governing both the opposite surfaces can be written in the form 
\begin{equation}
 \centering
 \mathcal{H}_{\text{s}}(\theta)=v_{\|}k_yS^x_\theta - v_1k_xS^y_\theta,
 \label{surf_ham}
\end{equation}
where $\mathbf{S}$ is the dispersing SU(2)  degree of freedom given by\cite{Zhang2012,Zhang2013} 
\begin{equation}
 \centering
 \mathbf{S_{\theta}}=\{\alpha_\theta\tilde{\sigma}^x-\beta_\theta\tau^z\otimes\tilde{\sigma}^z,\sigma^y,\alpha_\theta\tilde{\sigma}^z+\beta_\theta\tau^z\otimes\tilde{\sigma}^x\},
 \label{s_theta}
\end{equation}
with  $\alpha_{\theta}=v_z\cos\theta/v_3$, $\beta_{\theta}=v_{\|}\sin\theta/v_3$, $v_3=\sqrt{(v_z\cos\theta)^2+(v_{\|}\sin\theta)^2}$, and $\tilde{\sigma}^{x(z)} = \sigma^{x(z)}\cos\theta\pm\sigma^{z(x)}\sin\theta$. The above surface Hamiltonian features a pair of degenerate Dirac-like spectra given by the dispersion
\begin{equation}
E_{\mathbf{k},\pm}(\theta) = \pm\sqrt{v_1^2k_x^2+v_{\|}^2k_y^2}~.
\label{eq:spectrum}
\end{equation}
Note that the two Dirac cones correspond to the surface states on the two opposite surfaces, each surface has only one Dirac one, as these are strong TIs.

In the $\tau\otimes\sigma$ basis Eq.(\ref{surf_ham}) assumes a block diagonal form,
\begin{equation}
\mathcal{H}_{\text{s}}(\theta)=\begin{pmatrix}
                                    \bm{\sigma}.\bm{B}^+_{\bm{k}}(\theta) && 0 \\
                                    0 &&\bm{\sigma}.\bm{B}^-_{\bm{k}}(\theta)
                                   \end{pmatrix},
\label{eq:hammatrix}
\end{equation}
where the effective magnetic fields $\bm{B_k}^\pm$ are given by 
\begin{equation}
\begin{split}
\bm{B_k}^\pm=\{k_yv_\|(\alpha\cos\theta \pm & \beta\sin\theta), -k_xv_1,\\ & k_yv_\|(\mp\beta\cos\theta+\alpha\sin\theta\}.
\end{split}
\label{eq:bfield}
\end{equation}
Parametrizing these effective magnetic fields as 
\begin{equation}
 \bm{B}^\pm_{\bm{k}}=\vert\bm{B}_{\bm{k}}^\pm\vert\{\sin\theta_{\bm{k}}^\pm\cos\phi_{\bm{k}}^\pm,\sin\theta_{\bm{k}}^\pm\sin\phi_{\bm{k}}^\pm,\cos\theta_{\bm{k}}^\pm\},
\end{equation}
the two eigenstates coming from each of the two blocks can be written as
\begin{equation}
 \vert \chi^{\pm}_{E+}(\bm{k}) \rangle = \begin{pmatrix}
                    \cos\frac{\theta^{\pm}_{\bm{k}}}{2}\\
                    \sin\frac{\theta^{\pm}_{\bm{k}}}{2}e^{i\phi_{\bm{k}}^{\pm}}
                   \end{pmatrix}; ~~
 \vert \chi^{\pm}_{E-}(\bm{k}) \rangle = \begin{pmatrix}
                    -\sin\frac{\theta^{\pm}_{\bm{k}}}{2}\\
                    \cos\frac{\theta^{\pm}_{\bm{k}}}{2}e^{i\phi_{\bm{k}}^{\pm}}
                   \end{pmatrix}
\label{eq:eigenstates2}
\end{equation}
where $\vert \chi^{+}_{E+} \rangle$ and $\vert \chi^{+}_{E-} \rangle$ are the positive and negative energy eigenstates respectively of the first block, and $\vert \chi^{-}_{E+} \rangle$ and $\vert \chi^{-}_{E-} \rangle$ are of the second block of Eq.(\ref{eq:hammatrix}). The eigenstates of the Hamiltonian in Eq.(\ref{eq:hammatrix}) can then be trivially expressed as
\begin{equation}
\begin{split}
 \vert \psi^+_{E+} \rangle = \begin{pmatrix}
                \vert \chi^{+}_{E+}\rangle\\
                \bm{0}
               \end{pmatrix}; ~~&
 \vert \psi^-_{E+} \rangle = \begin{pmatrix}
		\bm{0}\\
                \vert \chi^{-}_{E+}\rangle
                \end{pmatrix}
\\
 \vert \psi^+_{E-} \rangle = \begin{pmatrix}
                \vert \chi^{+}_{E-}\rangle\\
                \bm{0}
               \end{pmatrix}; ~~&
 \vert \psi^-_{E-} \rangle = \begin{pmatrix}
		\bm{0}\\
                \vert \chi^{-}_{E-}\rangle
                \end{pmatrix}
\end{split}
\label{eq:eigenstates4}
\end{equation}
Note that the two-fold degeneracy of the spectra in Eq.~\ref{eq:spectrum} comes due to the fact that $\vert\bm{B}^+_{\bm{k}}\vert=\vert\bm{B}^-_{\bm{k}}\vert$.

The correct spinors describing the surface states on the two surfaces can be constructed by taking appropriate linear combinations of the states of the same energy. Obviously, the coefficients in the linear combinations can differ only by phase which we denote by $\phi_R$. This phase is fixed by demanding the state to be an eigenstate of $T_{\theta}^x$\cite{Zhang2012,Roy2014a} with eigenvalue $\pm 1$ (for the surface state of $\theta$ and $\pi+\theta$ respectively) where $\mathbf{T_{\theta}}$ represents the pseudospin degree of freedom given by
\begin{equation}
 \centering
 \mathbf{T_{\theta}}=\{\alpha_\theta\tau^x+\beta_\theta\tau^y\otimes\sigma^y,\, \alpha_\theta\tau^y-\beta_\theta\tau^x\otimes\sigma^y, \,\tau^z\}.
 \label{T_theta}
\end{equation}
Hence the very construction for surface state for any $\theta$ naturally involves imposing a complete $\mathbf{T}_{\theta}$ polarization on all the surface electron states independent of their momentum, making it a perfect analog of an ferromagnetic half metal where spins are fully polarized independent of momentum. We refer to these states respectively as the top and the bottom surface corresponding to a particular $\theta$. This leads to the expression of  $\phi_R$ as
\begin{equation}
e^{i\phi_R^{\text{t(b)}}}=+(-)\frac{\cos(\theta^+/2)}{\alpha\cos(\theta^-/2)-\beta\sin(\theta^-/2)e^{i\phi^-}}
\end{equation}
The resultant surface states are written as
\begin{equation}
 \vert \Psi_{c(v)}^{\text{t(b)}}(\bm{k}) \rangle =\frac{1}{\sqrt{2}}\left( \vert \psi_{E+(E-)}^+(\bm{k}) \rangle + e^{i \phi_R^{\text{t(b)}}} \vert \psi_{E+(E-)}^-(\bm{k}) \rangle \right),
\label{Psi}
\end{equation}
which ultimately leads to the surface states as 
\begin{equation}
\vert \Psi_{c(v)}^{\text{t(b)}}(\bm{k}) \rangle =\frac{1}{\sqrt{2}} \begin{pmatrix}
						~~~~~~~~ \vert\chi_{E+(E-)}^+(\bm{k})\rangle\\
						e^{i \phi_R^{\text{t(b)}}}\vert \chi_{E+(E-)}^-(\bm{k}) \rangle\\
					\end{pmatrix},
\label{eq:genpsi}
\end{equation}
where $\text{t(b)}$ in the superscript denotes the top(bottom) and $c(v)$ in the subscript denotes the conduction(valence) band.

\begin{figure}[t]
\centering
\includegraphics[width=0.5\columnwidth]{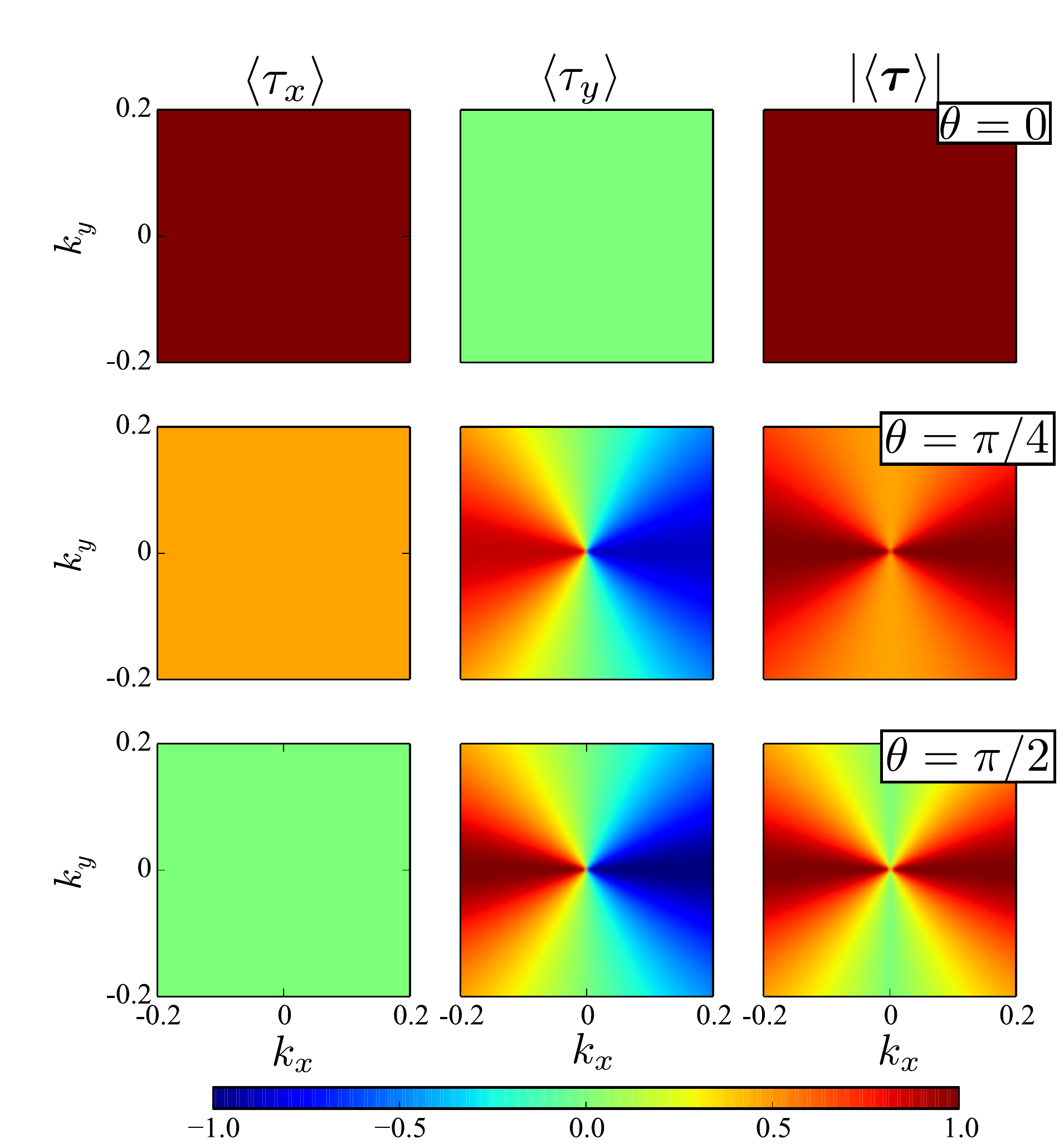}
\caption{The orbital texture for three different surfaces corresponding to $\theta=0,\pi/4, \text{and} \pi/2$ are plotted in the momentum space. The axes of the figures are $k_x$ and $k_y$ which are the local in-plane axes. It is important to note that for the $\theta=0$ surface, the orbital pseudospin shows a complete polarization in the pseudo $x$ direction, however as one moves towards a more oblique surface, this polarization begins to go down and at the same time, the polarization in the pseudo $y$ direction begins to pick up the texture. }
\label{fig:orbitaltexture}
\end{figure}

\subsection{Pseudospin textures}

The spin and orbital texture on the surface can be measured by just sandwiching the respective operators between the correct surface state spinors and calculating the expectation value. The spin expectation is given by $\langle\bm{\sigma}\rangle=(\langle\chi^+_{E+}\vert\bm{\sigma}\vert\chi^+_{E+}\rangle + \langle\chi^-_{E+}\vert\bm{\sigma}\vert\chi^-_{E+}\rangle)/2$ with the following expressions for the individual components
\begin{equation}
\begin{split}
\langle\sigma^x\rangle(\bm{k})&=\frac{k_yv_\|\alpha\cos\theta}{\sqrt{v_\|^2k_y^2 + v_1^2k_x^2}}\\
\langle\sigma^y\rangle(\bm{k})&=\frac{-k_xv_1}{\sqrt{v_\|^2k_y^2 + v_1^2k_x^2}}\\
\langle\sigma^z\rangle(\bm{k})&=\frac{k_yv_\|\alpha\sin\theta}{\sqrt{v_\|^2k_y^2 + v_1^2k_x^2}}
\end{split}
\end{equation}
By a similar calculation, the orbital pseudo spin texture in the conduction band can be also be formally expressed as 
\begin{equation}
\langle\bm{\tau}\rangle = \text{Re}[e^{i\phi_R} \bm{\tau}_{12}\langle\chi_{E+}^+\vert\chi_{E+}^-\rangle]
\end{equation}
Using the form of the eigenstates the component-wise expectation values can be obtained as\cite{Zhang2012} 
\begin{subequations}
\begin{equation}
\langle\tau_x\rangle =\cos\phi_R\sin\theta_{\bm{k}}=\frac{v_z\cos\theta}{v_3},
\label{eq:tx}
\end{equation}
\begin{equation}
\langle\tau_y\rangle = \sin\phi_R\sin\theta_{\bm{k}}=\frac{-v_zv_{\|}^2k_x\sin{\theta}}{v_3^2\sqrt{v_1^2k_x^2+v_{\|}^2k_y^2}},
\label{eq:ty}
\end{equation}
\begin{equation}
\langle\tau_z\rangle = 0.
\end{equation}
\end{subequations}
A few interesting features in the orbital pseudospin texture are of note. Firstly, as in the case of the spin texture, the magnitude does not stay constant over the Fermi surface, however, the expectation of $\langle\tau_x\rangle$ stays the same over the momentum space, though the value it takes depends on the surface taken. On the other hand, for any arbitrary surface not perpendicular to the crystal growth axis, $\langle\tau_y\rangle$ has a texture. Secondly, it is also intersting to note that the texture of the magnitude of the pseudospin in the momentum space mirrors that of the spin texture. Thirdly, the $\langle\tau_z\rangle$ is identically 0. This is a manifestation of the fact that on any surface, the contribution to the state coming from the two parity orbital sectors can differ only by a phase. This can also be seen from the form of the surface state spinor (Eq.(\ref{eq:genpsi})). Indeed in a low energy field theoretical description of the surface, this is what should be expected. These features are shown in Fig.(\ref{fig:orbitaltexture}). As in Fig.(\ref{fig:orbitaltexture}), the first row shows the surface which has the trivial orbital texture with the orbital pseudospin pointing unifomly in the $x$ direction in the pseudospin space.

The relative textures of two TI surfaces play a central role in the generalized magnetoresistance to tunneling current between them as we show in the next section.


\section{Generalized tunnel magnetoresistance \label{sec:gtmr}} 

In this section we sketch the calculation of the tunneling current between two TI surfaces from which the form of the generalized tunnel magnetoresistance naturally comes out. We consider a translation invariant tunnel Hamiltonian describing tunneling between two planar surfaces of TI with arbitrary $\theta$. The unperturbed Hamiltonian of the combined system (of the two TI's) can be written as 
\begin{equation}
\centering
 \mathcal{H}_0 = \int d\mathbf{k}~ ( E_1(\mathbf{k},\theta_1) c_{1,\mathbf{k}}^{\dag}c_{1,\mathbf{k}} + E_2(\mathbf{k},\theta_2) c_{2,\mathbf{k}}^{\dag}c_{2,\mathbf{k}} ),
 \label{unper_ham}
\end{equation}
where $c_{1/2,\mathbf{k}}$ is the fermionic annihilation operator for the two TI surfaces (with $\theta_1$ and $\theta_2$) and expression for $E_{1/2}$ is given in Eq.(\ref{eq:spectrum}). Translation invariance leads to a momentum resolved tunnel-Hamiltonian density given by
\begin{equation}
 \centering
 \mathcal{H}_{\text{T}} = {J}\int d\mathbf{k}~ \{z_\mathbf{k}c_{1,\mathbf{k}}^{\dag}c_{2,\mathbf{k}} + z_\mathbf{k}^\ast c_{2,\mathbf{k}}^{\dag}c_{1,\mathbf{k}}\},
 \label{tunnel_ham}
\end{equation}
where $z_\mathbf{k}$ is the overlap between the two surface state wave functions at momentum $k$. 

\noindent
The resultant current at time $t$ can be written as
\begin{eqnarray}
 \langle \hat{I}(t)\rangle  &=& \langle GS(t)|\hat{I}(t)|GS(t)\rangle  \nonumber\\         
                            &=& \langle g \vert \int_{-\infty}^t dt'~ [\mathcal{H}_{\text{T}},e^{i\mathcal{H}_0 t/{\hbar}} \hat{I} e^{-i\mathcal{H}_0 t/{\hbar}}] \vert g \rangle,
\end{eqnarray}
where $|GS(t)\rangle$ and $\vert g\rangle$ are respectively the time-dependent and time-independent ground states of the combined system and the last line in the above equation is obtained keeping the leading term in $J$ in the Dyson series of the evolution operator\cite{bruus2004many}. Using the algebra of the fermionic operators and time translation invariance we arrive at
\begin{widetext}
\begin{equation}
 \langle \hat{I}(t \rightarrow \infty)\rangle := \langle I\rangle = \frac{J^2e}{\hbar} \langle g \vert \int_{-\infty}^{\infty} dt \int d\mathbf{k} \vert z_{\mathbf{k}}\vert^2 [ b^{\dagger}_{\mathbf{k}}(t) b_{\mathbf{k}}(0) a_{\mathbf{k}}(t) a^{\dagger}_{\mathbf{k}}(0)  -   b_{\mathbf{k}}(0) b^{\dagger}_{\mathbf{k}}(t) a^{\dagger}_{\mathbf{k}}(0) a_{\mathbf{k}}(t)         ] \vert g \rangle,
\end{equation}
which finally leads to the following expression of the tunneling current density given by
\begin{equation}
 \centering
 \langle I\rangle = \frac{J^2e}{\hbar}\int d\mathbf{k}\;\vert z_{\mathbf{k}}\vert^2 \Delta n_\text{F}(\mathbf{k},\mu_1,\mu_2) \times \int dE [\delta(E_1(\mathbf{k})-E)\delta(E_2(\mathbf{k})-E)].
\end{equation}
\end{widetext}
Here $ \Delta n_\text{F}(\mathbf{k},\mu_1,\mu_2)=n_{\text{F}}(E_1(\mathbf{k}),\mu_1)-n_{\text{F}}(E_2(\mathbf{k}),\mu_2)$ is the difference in the Fermi functions of the two surfaces and
\begin{equation}
\vert z_{\mathbf{k}}\vert^2 = [1+\langle\bm{\sigma}\rangle_1\cdot\langle\bm{\sigma}\rangle_2+\langle\bm{\tau}\rangle_1\cdot\langle\bm{\tau}\rangle_2 +\displaystyle\sum_{i,j}\langle\tau_i\otimes\sigma_j\rangle_1\langle\tau_i\otimes\sigma_j\rangle_2]/4.
\label{eq:zk2}
\end{equation}

Eq.\ref{eq:zk2} is the expression of the generalized tunnel magnetoresistance for tunneling between two TI surfaces and is one of the main results of the work. This decomposition is both nontrivial and instructive as it allows for interpretation of the total tunneling current in terms of individual responses from  $\bm{\sigma}$,~$\bm{\tau}$ sectors and a correlation term between them. The surface states being eigenstates of $T_{\theta}^x=\alpha_\theta\tau^x+\beta_\theta\tau^y\otimes\sigma^y$ lead to a momentum independent part in $\vert z_{\mathbf{k}}\vert^2$ given by $\vert z_f \vert^2 \equiv\langle\tau_x\rangle_1\langle\tau_x\rangle_2 + \langle\tau_y\otimes\sigma_y\rangle_1\langle\tau_y\otimes\sigma_y\rangle_2$, which is representative of the overlap between the frozen sectors of the respective surfaces which does not depend on the momentum. These frozen sectors are crucial towards identifying the half-metallic degree of freedom on the TI surfaces as we shown in the next section.


\section{Identifying the half-metallic degree of freedom \label{sec:halfmetal}} 

In this section we explicitly identify the half-metallic degree of freedom via tunneling current measurements. We specifically choose to work with junctions of $\theta=0$ and $\pi$ surfaces, as $\mathbf{S_{\theta}}$ and $\mathbf{T_\theta}$ reduce to the spin $\bm{\sigma}$ and the orbital pseudospin $\bm{\tau}$ respectively. Consequently the expression for the general tunnel magnetoresistance \eqref{eq:zk2} simplifies to 
\begin{equation}
\vert z_{\mathbf{k}}\vert^2 = 
\left(\frac{1+\langle\bm{\sigma}\rangle_1(\mathbf{k})\cdot\langle\bm{\sigma}\rangle_2(\mathbf{k})}{2}\right)\left(\frac{1+\langle\bm{\tau}\rangle_1\cdot\langle\bm{\tau}\rangle_2}{2}\right).
\label{eq:zk2top}
\end{equation}

The first term in Eq.\ref{eq:zk2top} represents the momentum dependent overlap of spin orientations of the  spin-momentum locked states on the two surfaces. The second term is of more interest here as it encodes the momentum-independent overlap of the pseudospin sectors of the surface states and the robust half-metallic beahvior can arise from this sector. Now, we know that $\bm\tau$ points along positive and negative x-axis for $\theta =0$ and  $\theta =\pi$ surfaces for a given TI crystal. Hence for a junction between $\theta =0$ surface of one TI with the $\theta =\pi$ surface of another TI, the first term in Eq.(\ref{eq:zk2top}) vanishes identically ($\langle\bm{\tau}\rangle_1\cdot\langle\bm{\tau}\rangle_2=-1$) leading to a perfect orthogonality in the pseudospin sector of the wavefunctions which for the $\theta=0,\pi$ surfaces is the pure orbital pseudospin sector. This directly stems from the half metallic behavior discussed above (see Junction-A in  Fig.(\ref{fig:setup})). 

Next we note that, if both the participating surfaces at the junction are $\theta=0$ surfaces for their respective crystals (see Junction-B in Fig.(\ref{fig:setup})), the orbital orthogonality is lifted ($\langle\bm{\tau}\rangle_1\cdot\langle\bm{\tau}\rangle_2=1$). But,  even if we start with two identical crystals with their crystal growth axis being parallel (taken as z-axis) to each other, still we have to rotate the crystal growth axis of one of the crystals with respect to the other by $\pi$ so that the $\theta=0$ surfaces of the two crystals could face each other hence forming the junction. A careful observation immediately reveals that the $\pi$ rotation of the crystal growth axis leads to a perfect orthogonality between the spins ($\bm\sigma$) of the electron states on the two surfaces for a given momentum. Hence, though the orbital orthogonality gets lifted in the Junction-B, an orthogonality spin sector ($\langle\bm{\sigma}\rangle_1(\mathbf{k})\cdot\langle\bm{\sigma}\rangle_2(\mathbf{k})=-1$) arises leading to zero tunneling current.  So, we have zero tunneling current for both Junction-A and -B. This fact makes the possibility of distinguishing the parallel (Junction-B) from anti-parallel (Junction-A) orbital pseudospin configuration obscure as the absence of the orbital orthogonality in Junction-B is masked by the orthogonality in the spin sector.

The half-metallic behavior manifests itself as the orbital orthogonality in Junction-A, hence to contrast its presence and absence in Junction-A and -B respectively, the spin ($\bm\sigma$) orthogonality has to be lifted. This can be achieved by strong doping which is large enough to push the Fermi surface into the hexagonally warped regime of the energy spectrum\cite{Fu2009}. Next, we demonstrate how this can be used to lift the spin orthogonality.

\begin{figure}[t]
\centering
\subfloat{\includegraphics[width=0.3\columnwidth]{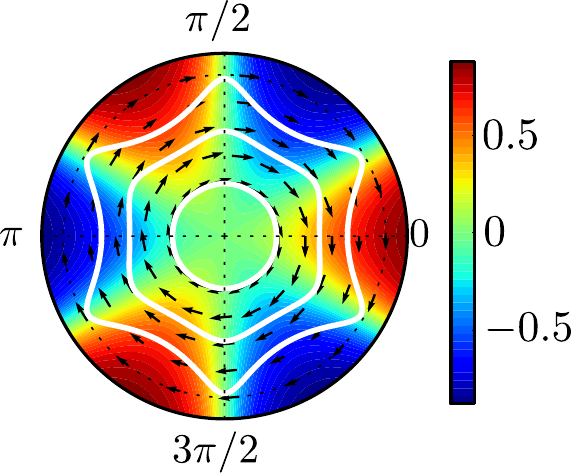}}
\subfloat{\includegraphics[width=0.4\columnwidth]{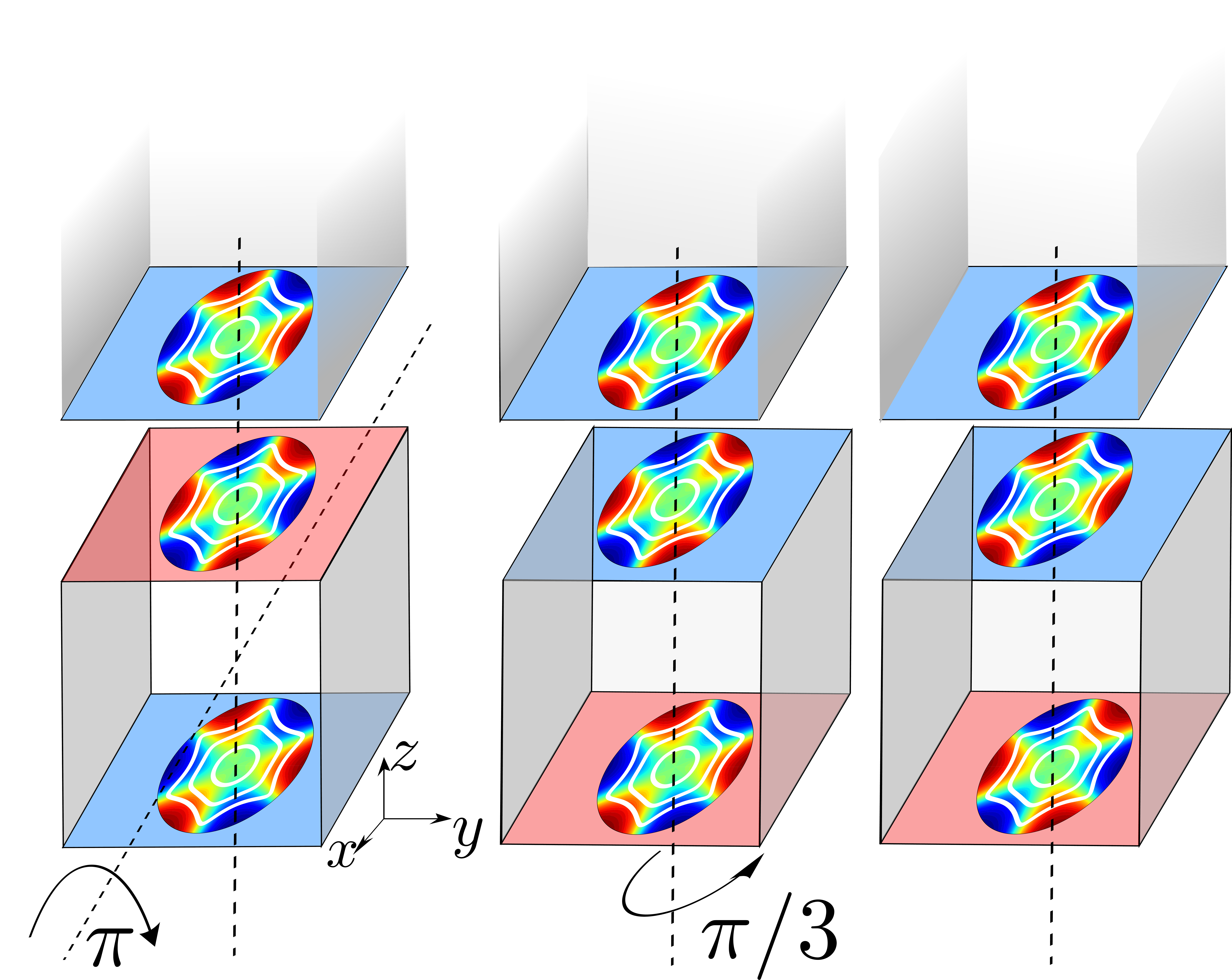}} 
\caption{{\bf Left}: The color density plot shows the out of plane magnetization ($\langle\sigma^z\rangle$) and the vectors show the in-plane $x$ and $y$ components for the surface states in the presence of warping.
{\bf Right}: The figure shows that how the spin orthogonality stays intact even after the orbital pseudo spin orthogonality is lifted by rotating the sample by $\pi$ such that both the surfaces participating in the junction have their orbital pseudo spin pointing uniformly in the same direction. The red and blue shades in the surface refer to the states having $\langle\tau_x\rangle=\pm1$ respectively and the vectors show the spin texture on the Fermi surface.}
\label{fig:warpspintexture}
\end{figure}

The low energy Hamiltonian for the TI surfaces to next higher order after linear is a term cubic in momentum which couples to the out-of-plane magnetization. The Hamiltonian with the inclusion of the warping term is given by
\begin{equation}
\mathcal{H}_{\text{W}} = (\bm{\sigma}\times\mathbf{k})_z+\frac{\lambda}{2}(k_+^3+k_-^3)\sigma^z.
\end{equation}
Such warping term induces a 3-fold symmetric pattern in the out-of-plane spin polarization $\langle\sigma^z\rangle$, with the Fermi surface breaking up into six symmetric regions of alternating positive and negative $\langle\sigma^z\rangle$ as shown in the left panel of Fig.(\ref{fig:warpspintexture}). The momenta with polar angle $\theta_{\mathbf{k}^*} = n\pi/3,$ where $n=0,1\dots5$, have maximal $\langle\sigma^z\rangle$ with its sign given by $(-1)^n$.

\begin{figure}
\includegraphics[width=0.5\columnwidth]{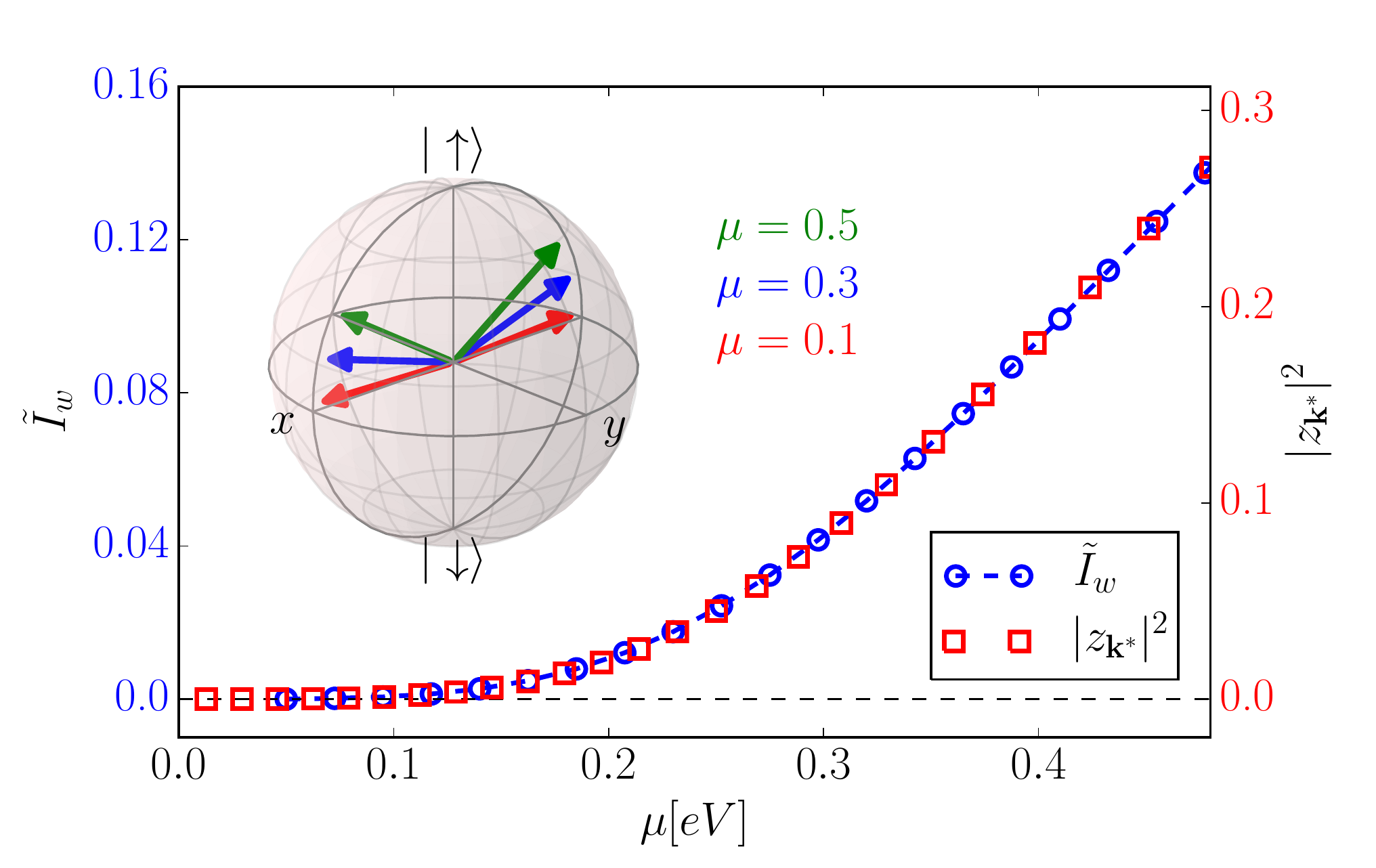}
\caption{The current, $\tilde{I}_w$, (normalized by the current for $\vert z_{\mathbf{k}}\vert^2=1~\forall \mathbf{k}$) for Junction-B and $\vert z_{\mathbf{k}^*}\vert^2$ ($\theta_{\mathbf{k}^*}=0$) is plotted against the average chemical potential $\mu$.  The plot shows that they are linearly proportional to each other. The inset shows the behavior of the spins on the $\theta_{\mathbf{k^*}}=0$ line for the two surfaces on the Bloch sphere for three different $\mu$.}
\label{fig:warp}
\end{figure}

The presence of the out-of-plane magnetization allows us to rotate one of the samples so as to lift the spin orthogonality as shown in Fig.(\ref{fig:warpspintexture}) right panel, which allows us to contrast the presence and absence of the orbital pseudo spin orthogonality by observing zero or finite current respectively in the junction. As shown in Fig.(\ref{fig:warpspintexture}), the top and bottom surfaces with opposite orbital pseudo spin $\langle\tau_x\rangle=\pm1$ (denoted by the blue and red shades) have opposite spin textures even in the out of plane component. A rotation about $x$-axis by $\pi$ lifts the orbital pseudospin orthogonality but still the spin orthogonality persists. This is where the 6-fold symmetry in the out-of-plane spin texture is exploited and it can be seen that a rotation of the second sample by $\pi/3$ about the crystal growth axis, leads to a configuration where the states of the two surfaces with same momentum have identical spin components out-of-plane even though the in-plane components continue to be opposite. This is enough to create a finite overlap between the states in the spin sector and lift the spin orthogonality.
In this situation, increasing the doping increases the warping effect leading to an increase in the overlap of the spin polarization of the two surfaces.
Hence the tunneling current grows as a function of doping (see Fig.(\ref{fig:warp})). The fact that the finite tunneling current comes only from the non-zero overlap of the out-of-plane spin components can be verified by noting that the overlap of the spin components at momenta $\mathbf{k}^*=n\pi/3$ is exactly proportional to the total current as shown in Fig.\ref{fig:warp}. For obtaining a geometric interpretation of the effect, maximally overlapping spins of the two surfaces state at momenta corresponding to states with maximal out of plane polarization (which contribute maximally to the tunneling current) is schematically shown on the Bloch sphere as a function of doping in inset of  Fig.(\ref{fig:warp}).

So, we have explicitly identified that for $\theta=0,\pi$ surfaces the orbital pseudospin $(\bm{\tau})$ degree of freedom constitutes a half-metallic degree of freedom owing to complete polarization which is independent of momentum. We further showed that how the effect of warping can be exploited to demonstrate this half-metallic behavior via spin-valve-like effect, but with the orbital pseudospin. 
For arbitrary surfaces $\theta\ne0,\pi$ the half-metallic degree of freedom is $\mathbf{T}_\theta$ degree of freedom and similar calculations can be done to demonstrate it. 
Note that, since for the $\theta=0,\pi$ surfaces, the orbital pseudospin degree of freedom is completely frozen and has no dependence on the momentum ($\tau^x(\mathbf{k})=\pm 1$), twisting the two samples in the junction about the crystal growth axis does not affect the orthogonality arising from the half-metallic behavior. On the other hand, for $\theta\ne0,\pi$ surfaces, though the $\mathbf{T}_\theta$ degree is frozen ($T^x_\theta(\mathbf{k})=\pm 1$) leading to the half-metallic behavior but in junctions of the form Junction-B the tunneling current could be affected by the anisotropy of the Fermi surface. Since the Fermi surfaces would be elliptical for $\theta\ne0,\pi$ surfaces, twisting the two surfaces in the junction relative to each other can lead energy and momentum being simultaneously conserved only at discrete points leading to a different value of the tunneling current.
For junctions of surfaces corresponding to different $\theta$s the different shapes of Fermi surfaces \eqref{eq:spectrum} leads to further features in the tunneling current which we discuss in the next section.
%

%
\section{Tunneling current between arbitrary surfaces \label{sec:arbitrary}}
\begin{figure*}
\centering
\subfloat{\includegraphics[width=0.33\linewidth]{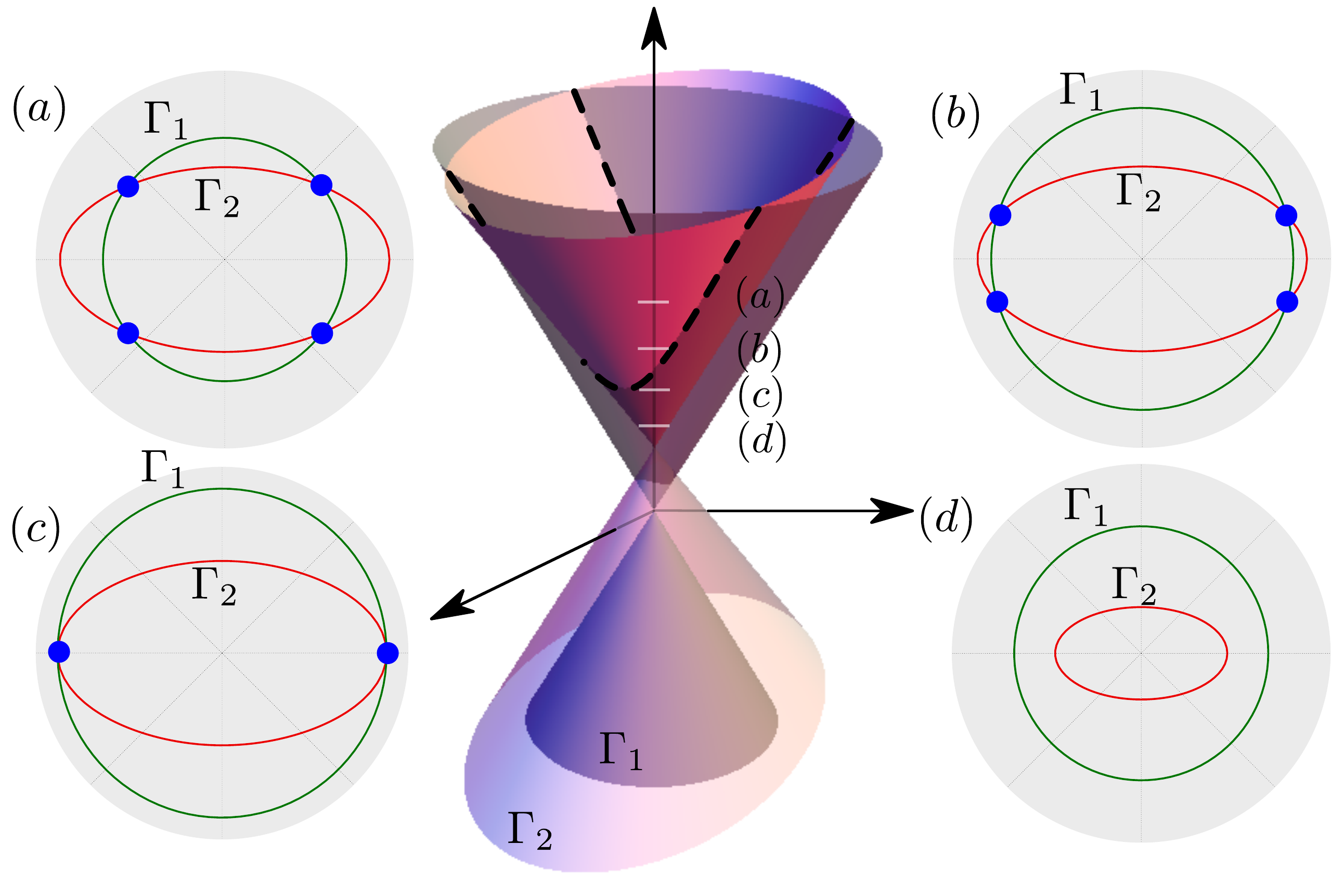}}
\subfloat{\includegraphics[width=0.33\linewidth]{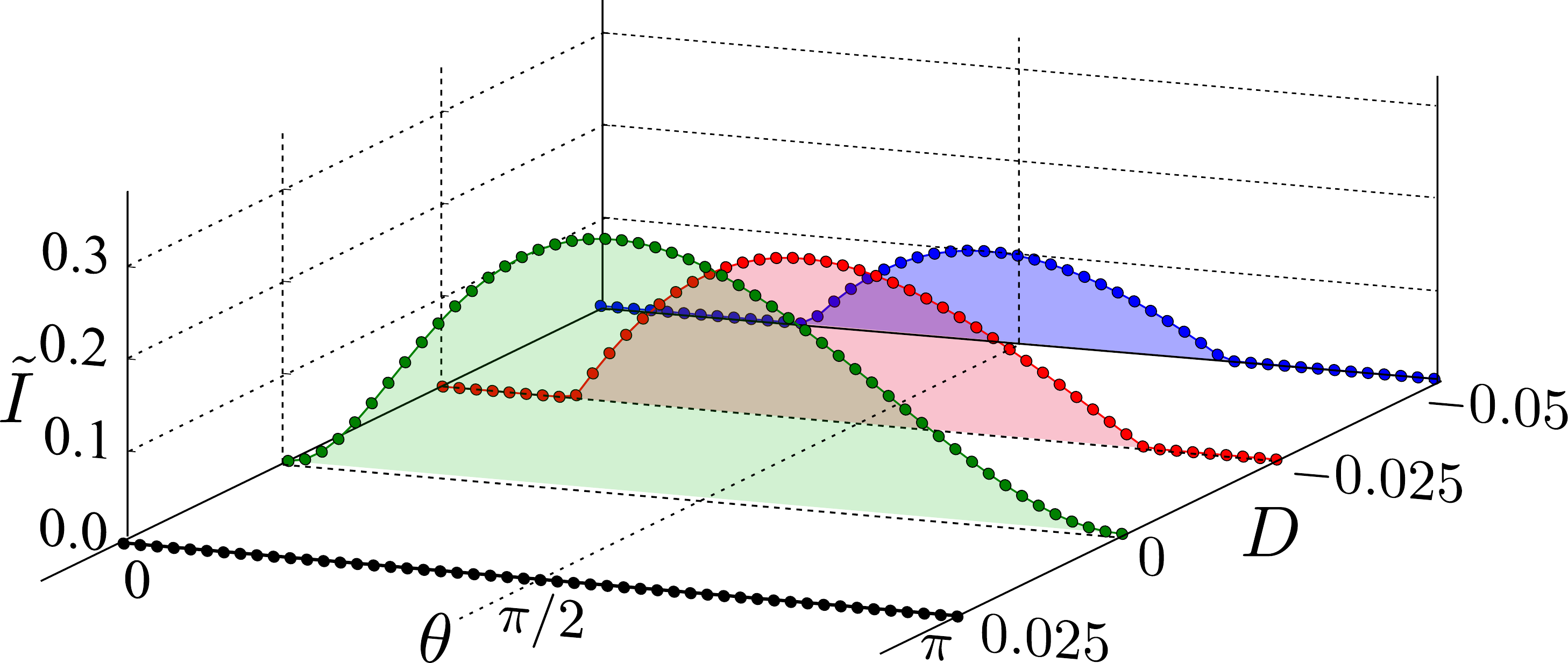}}
\subfloat{\includegraphics[width=0.33\linewidth]{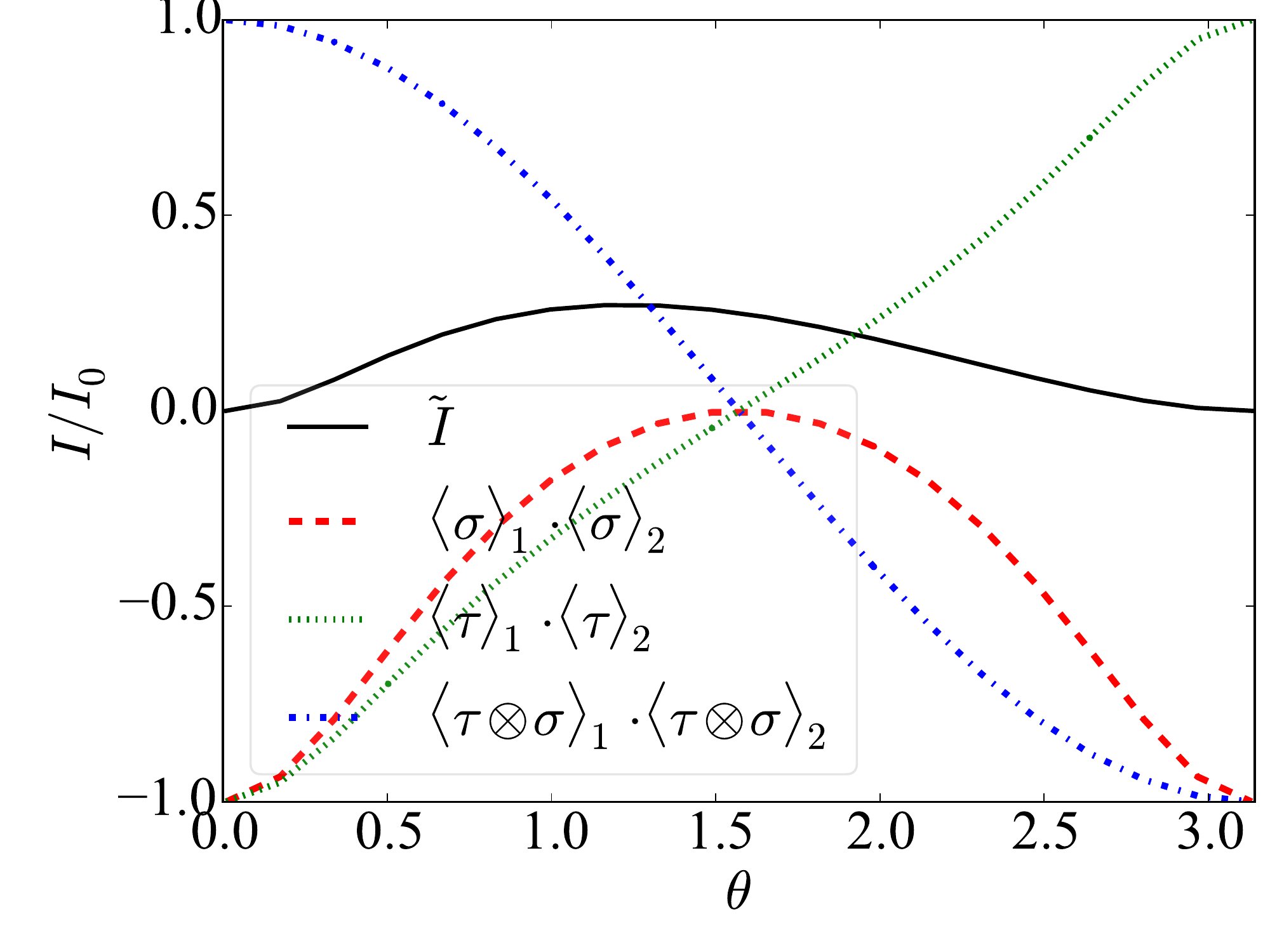}} 
\caption{{\bf Left}: The two spectrums, one conical and one ellitpical are shown on top of each other and the dotted line shows the locus of the points with equal energy and momentum. The subplots show the Fermi surfaces at four different energies, which show the presence of two or four, or the absence of Fermi surface intersections. {\bf Center}: The normalized current ($\tilde{I}$) is plotted as a function of $\theta$ for four different doping. To lift the no-match condition at higher negative dopings of the circular Fermi surface, one needs to deviate more from $\theta=0$ as eccentricity of the elliptical Fermi surface increases with $\theta$ till $\pi/2$ and then goes down. {\bf Right}:The asymmetry in the current about $\theta=\pi/2$ is understood by looking at the contribution of the different sectors to the current.}
\label{fig:spectrum}
\end{figure*}
Tunneling between the two opposite TI surfaces corresponding to the same $\theta$ can always be suppressed due to the orthogonality in the half-metallic sector ($\mathbf{T}_\theta$) as was shown in Sec.\ref{sec:halfmetal}. However, tunneling between two surfaces corresponding to different $\theta$ can also be suppressed for an entirely different reason, namely the failure to conserve energy and momentum simultaneously, which we analyse in this section.

The Fermi surface of the surface states is actually an ellipse whose eccentricity depends on the angle $\theta$ \eqref{eq:spectrum}. Conserving energy and momentum simultaneously would mean that the Fermi surfaces of the two TIs should intersect in the momentum space within the bias window. Failure to meet this condition leads to a zero current and we call it a {\it{no-match condition}}. The presence or absence of this condition is dictated by the angles the two surfaces make with their crystal growth axes and their respective dopings. For simplicity we take one of the surfaces of the junction to be the bottom surface perpendicular to the crystal growth axis with a circular Fermi surface doped by an amount $D$ with repect to the second surface. If the second surface of the junction is allowed to be arbitrary (characterized by $\theta$), then for a given energy $E$, the points of intersection of the Fermi surfaces denoted by $\mathbf{k}^\ast=(k^\ast,\theta_{k^\ast})$ are parametrized by the doping fraction $f$ ($\equiv D/E$) and $\theta$ as
\begin{equation}
{\theta_{k^*}}(\xi)= 
\begin{dcases}
    \pi/2,3\pi/2 & \text{if } \xi=-1\\
    0,\pi                        & \text{if } \xi=1\\
    \theta_{k1}^*,\theta_{k1}^*+\pi,\theta_{k2}^*,\theta_{k2}^*+\pi  & \text{if } \vert\xi\vert<1,
\end{dcases}
\label{eq:solns}
\end{equation}
where $\xi=2 [v_3/(1-f)]^2 -[(v_z^2+v_3^2)/(v_z^2-v_3^2)]$; $\theta_{k1}^*=\cos^{-1}(\sqrt{(1+f)/2})$ and $\theta_{k2}^*=\cos^{-1}(-\sqrt{(1+f)/2})$, with $k^\ast = (E-D)/\hbar v_\|$. For $\vert\xi\vert>1$, there exists no such intersections.

This is illustrated in the left panel of Fig.\ref{fig:spectrum} by overlaying the two spectrums and showing the presence or absence of intersections at different energies. The energy spectrum in Eq.(\ref{eq:spectrum}) shows that for the elliptical Fermi surfaces, the semi-minor axes ($v_\|^{-1}$) is equal to the radius of the circular Fermi surface which means they always touch at two points for $D=0$, however when one of the samples is doped, the Fermi surfaces intersect only above a threshold energy. Using the locations of Fermi surface intersections from Eq.(\ref{eq:solns}) and the overlap of the spinors at these intersections from Eq.(\ref{eq:zk2}), the current expectation is evaluated and normalized by a value that would have been present had the two spinors been parallel in the SU(4) space to define a dimensionless quantity ($\tilde{I}$) plotted in the center panel of Fig.\ref{fig:spectrum}. The dimensionless current $\tilde{I}$ is plotted as a function of $\theta$ for different dopings of the first sample (with a circular Fermi surface) keeping the center of the tiny bias window fixed at a positive chemical potential well away from the neutrality point. The current is always zero for $D>0$, due to the aforementioned {\it{no-match condition}}, as in this case, the radius of the circle of the Fermi surface for a given energy $E$ shrinks by $D/\hbar v_\|$ and there are no intersections with the elliptical Fermi surface of the other surface. However for $D<0$, the circular Fermi surface's radius increases by $\vert D\vert/\hbar v_\|$, and as one deviates from $\theta=0$ the eccentricity of the Fermi surface increases leading to intersections. The eccentricity of the Fermi surface being symmetric about $\theta=\pi/2$ increases with $\theta$ till $\pi/2$, so for higher negative dopings of the circular Fermi surface, one needs to go to higher deviations from $\theta=0 \text{ or } \pi$ to lift the {\it{no-match condition}} and get a finite current. This is observed the fact that range of $\theta$ around $\pi/2$ over which there is finite current, shrinks for higher negative dopings. Also, note that for $D<0$ the tunneling current is aymmetric about $\theta=\pi/2$. This can be understood from looking at the contributions of individual terms to the generalized tunnel magnetoresistance separately as shown in the right panel of Fig.\ref{fig:spectrum}.
%


\section{Robustness of the half-metal to surface potentials \label{sec:surfacepotentials}} 

The half-metallic behavior of the surface state of 3D TI and its diagnostics via the orbital orthogonalities solely rely upon the fact that the surface states on the two opposite planar surfaces of a TI sample are orthogonal eigenstates of the $T_\theta^x$ operator. However, a realistic scenario would naturally involve influence of surface potentials\cite{Zhang2012,RD2016} which need not keep this orthogonality intact. The surface potentials considered preserve time reversal symmetry(TRS), else the spectrum of the surface states itself can become gapped due to its influence and that will not be of interest for our study. These potentials not only change the nature of the surface states leading to a change in expectation values of the spin and the orbital pseudospin but they can also shift the position of the Dirac point in energy leading to doping effects\cite{Zhang2012}. We systematically study the influence of these potentials on the pseudospin-valve effect for the $\theta=0,\pi$ surfaces. 

The various classes of potentials allowed by TRS can be listed as $\mathds{I}_4$~,~($\mathbf{\Delta}_T\cdot \mathbf{T}_\theta)$ and $( \mathbf{\Delta}_{S}\cdot\mathbf{S}_\theta)T^y$ which for the $\theta=0,\pi$ surfaces look like $\mathds{I}_4$~,~($\mathds{I}_2 \otimes\mathbf{\Delta}_{\tau}\cdot\bm{\tau})$ and $(\mathbf{\Delta}_\sigma\cdot\bm{\sigma})\,\otimes\tau^y$ where $\mathbf{\Delta}_\tau$ is a vector that lives only in the $x$-$z$ plane in the $\tau$-space and $\mathbf{\Delta}_{\sigma}$ lives in the full 3D $\sigma$-space. It is important to note that the spin and orbital pseudospin degree of freedom are always decoupled for the $\theta=0,\pi$ surfaces even in the presence of the surface potentials, so one can analyze their effect on the spin and orbital pseudospin sector separately. We first consider a surface potential of the form $\Delta \,\delta(z)\,\mathds{I}_4$ where the surface is taken to be at the $z=0$ plane. This potential results in change of the orbital pseudospin texture of the surface states, but keeps the spin texture intact. As a result, the orbital orthogonality gets lifted. However we find the overlap in the orbital pseudospin sector ($(1+\langle\bm{\tau}\rangle_1\cdot\langle\bm{\tau}\rangle_2)/2$) is  proportional to $m^2\Delta_0^2(m^2-\Delta_0^2)^2/(m^2+\Delta_0^2)^4$ where $\Delta=(\Delta_0/m) (2 v_z)$ and  $m$ is the bulk band gap of the TI. Note that the overlap tends to zero both in the $m/\Delta_0\gg1$ and $m/\Delta_0\ll1$ limit and has significant contribution only in window around $m/\Delta_0\approx1/2$. Hence for a junction with weak  surface potential strength ($\Delta$) will not effect the orbital orthogonalitys significantly. Next we consider the potential $\Delta\,\delta(z)\,\mathds{I}_2\otimes\tau^z$. Such a potential leaves both the spin and orbital textures of the surface states intact.  Hence it does not effect the orbital orthogonality. Potentials of the form $\tau^x\otimes\mathds{I}_2$ change the orbital pseudospin texture of the surface states and hence it potentially can lift the orbital orthogonality. However, we find that the effect of this surface potential is the same on both the surfaces for Junction-A as far as the orbital pseudospin is concerned and hence for a symmetric junction (same strength of surface potential on both the surfaces) such a potential should not have any effect. Also, though this potential does not affect the spin texture of the surface states but it does lead to lifting of the spin orthogonality owing to shift in the Dirac spectrum in opposite direction on the two surfaces. Finally we consider potentials of the form $\mathbf{\Delta}_{\sigma}\cdot\bm{\sigma}\otimes\tau^y$ which keep the orbital pseudospin texture intact, however they do affect the spin texture. Hence they preserve the orbital orthogonality expected to be observed in Junction-A, however in Junction-B since the orbital orthogonality is anyway absent, and these potentials destroy the spin orthogonality too and hence  one can expect a finite conductance even in the absence of either an in-plane magnetic field or a doping to the warped regime of the spectrum. To summarize, we find that among all TRS preserving surface potentials, only $\tau^x\otimes\mathds{I}_2$ and $\mathds{I}_4$ affect the orbital pseudospin texture of the surface states. But even for these two potential we pointed out situations where the orbital orthogonality will not get destroyed. A summary of effect of these potentials on the orthogonalitys can be found in Table.\ref{tab:surfacepot}.\\
\begin{table}[!t]
\begin{tabular}{|c|c|c|c|c|c|c|}
 \cline{1-7}
 \multirow{2}{*}{}& \multirow{2}{*}{$0$} & \multirow{2}{*}{$\mathds{I}_4$} & \multicolumn{2}{c|}{$\mathds{I}_2\otimes\bm{\Delta}_{\tau}\cdot\bm{\tau}$} &\multicolumn{2}{c|}{$\bm{\Delta}_{\sigma}\cdot\bm{\sigma}\otimes\tau^y$}\\
 \cline{4-7}
 & & & $\bm{\Delta}^{(1)}_{\tau}=\bm{\Delta}^{(2)}_{\tau}$ &$\bm{\Delta}^{(1)}_{\tau}\neq\bm{\Delta}^{(2)}_{\tau}$ &$\bm{\Delta}^{(1)}_{\sigma}=\bm{\Delta}^{(2)}_{\sigma}$ &$\bm{\Delta}^{(1)}_{\sigma}\neq\bm{\Delta}^{(2)}_{\sigma}$\\
 \hline \hline
 A & \cmark & \xmark & \cmark & \xmark & \cmark & \cmark\\
 \hline
 B & \xmark & \xmark & \xmark & \xmark & \xmark & \xmark\\
 \hline
\end{tabular}
 \caption{The table shows the effect of the different surface potentials on the orbital ($\tau$) orthogonalities on both Junctions-A and Junction-B for the $\theta=0,\pi$ surfaces. The column corresponding to $0$ refers to case of no surface potentials. A \cmark (\xmark) denotes the presence (absence) of the orthogonality.}
 \label{tab:surfacepot}
\end{table}

\section{Conclusion \label{sec:conclusions}} 

We show that the surface states of 3D TI exhibit complete polarization in an emergent pseudospin (SU(2)) degree of freedom  and  hence behave like a half-metal. We find that this half-metallic behavior is robust against most of the surface potentials that preserve TRS. Since this half-metallic behavior leaves its signatures in transport properties between two surface states, transport probes to reveal this are also proposed. We showed that the half-metallic behavior can lead to a pseudospin-valve-like effect in the tunneling current between two TI surfaces, which can be used as a diagnostic for the half-metallic behavior. Our study indicates that weak disorder (which could get induced due to the roughness in surface potentials) can not only affect transport due to scattering of electron due to its coupling with the spin degree of freedom of the disorder potential (which is the one usually considered for study \cite{mirlin}) but
also due to its coupling with the frozen pseudospin degree of freedom of the disorder potential. Presence of such disorder may not be seen in the traditional methods used to probe surface states which couple only to the spin texture of the surface states such as the spin polarized ARPES and STM. 
Hence junctions of the type proposed in this letter can play a crucial role in understanding these pseudospin degree of freedom. It should be further noted that all the results obtained for the $\theta=0,\pi$ surfaces can be extended in a straightforward manner to arbitrary surfaces using Eq.\eqref{eq:zk2} or Eq.\eqref{eq:zk2top} with $\sigma\rightarrow S_\theta$ and $\tau\rightarrow T_\theta$. For arbitary surfaces, the effect of the elliptic shape of the Fermi surface was also analyzed.

To conclude, some remarks are in order about the applications of our results to topological insulator materials belonging to classes different from $\mathrm{Bi}_2\mathrm{Se}_3$. A notable example is the topological crystalline insulator in the SnTe material class\cite{HLLDBF2012,zeljkovic2014mapping} which posses even number of Dirac cones on the surfaces. The surface states in such materials are not only spin-polarized but also have another  degree of freedom namely the ``mirror eigenvalue''\cite{HLLDBF2012} whose topological robustness arises from the reflection symmetries of the crystal. We expect that the interplay of these different degrees of freedom can lead to further rich physics in the context of mesoscopic transport which is left here as a subject of future research.


\section*{Acknowledgments} 

The authors are deeply indebted to  E. J. Mele for the insightful and illuminating discussions regarding the incorporation of surface potentials.

\bibliography{references}
\end{document}